\documentstyle[aps,prl]{revtex}
\begin{document}

\draft
\twocolumn[\hsize\textwidth\columnwidth\hsize\csname
           @twocolumnfalse\endcsname

\preprint{DAMTP-R97/42}
\title{Chaos in Quantum Cosmology}
\author{ N.~J.~Cornish and E.~P.~S.~Shellard} 
\address{DAMTP. University of Cambridge, Silver Street, Cambridge CB3
9EW, UK}
\maketitle
\widetext
\begin{abstract}
Most of the foundational work on quantum cosmology employs a simple
minisuperspace model describing a Friedmann-Robertson-Walker universe
containing a massive scalar field. We show that the classical limit of this
model exhibits deterministic chaos and explore some of the consequences
for the quantum theory. In particular, the breakdown of the WKB
approximation calls into question many of the standard results in
quantum cosmology.
\end{abstract}
]
\narrowtext
\begin{picture}(0,0)
\put(400,170){{\large DAMTP-R97/42}}
\end{picture} 
\vspace*{-0.2 in}

It has been suggested by Hawking\cite{hawk1} that the isotropy and
homogeneity of the universe is a natural consequence of the
Hartle-Hawking\cite{hh} ``no-boundary'' boundary conditions for the
quantum state of the early universe. Testing this idea has proved
difficult since a realistic model of the quantum state or wave function
of the universe must describe the entire geometry and matter content
of the universe. In order to make progress, a number of
simple models have been introduced. The most extensively studied model
employs a homogeneous and isotropic Friedmann-Robertson-Walker (FRW)
spacetime with a minimally coupled massive scalar
field. Linde\cite{linde} showed that for certain initial conditions
this model could lead to exponential expansion of the universe and a
successful realisation of the inflationary paradigm.

The FRW-scalar field model has been the work-horse of quantum
cosmology. It is the model upon which most of the successes of quantum
cosmology are based\cite{hawk2,alex,hawk3}, including Hawking's claim
that the no-boundary proposal predicts inflation\cite{hawk2}. In this
letter we describe a problem that complicates the interpretation of these
results. We show that the classical trajectories of the model exhibit
a form of deterministic chaos known as chaotic
scattering\cite{ott}. This is a cause for concern since the quantum to
classical transition is qualitatively different for chaotic and
non-chaotic systems. The problem is particularly pronounced in quantum
cosmology since the theory is manifestly
semi-classical\cite{hh,linde}, and since the wavefunction of the
universe is interpreted using the WKB approximation\cite{hawk2,hp,alex2}.
As first pointed out by Einstein\cite{al}, the WKB approximation
fails if the classical evolution is chaotic.

Considering how much work has been done studying such a simple model,
it may seem surprising that the chaotic behaviour went unnoticed.
In fact, the fingerprints of chaos can be found in some of the early
literature\cite{page,shell,kiefer}, but it is only now, with an understanding
of chaotic scattering, that we can recognise them as such.

The model in question describes a closed FRW universe with Lorentzian metric
\begin{equation}\label{lorentz}
ds^2=-N^2(t) dt^2 + a^2(t) d\Omega_{(3)}^2 \, ,
\end{equation}
minimally coupled to a scalar field $\phi$. A fuller description of the
model can be found in Ref.~\cite{shell}. The minisuperspace
action is given by $S=\frac{1}{2}\int  (N H / a)\, dt$ where
\begin{equation}
H= -\left({a \over N}
{d a \over dt}\right)^2 + \left({a^2 \over N} {d \phi \over
dt}\right)^2 -V(a,\phi) 
\end{equation}
is the classical Hamiltonian and $V=-a^2+m^2a^4\phi^2$ is the
minisuperspace potential. 
The classical equations of motion that follow from varying $\phi$ and $a$ are:
\begin{eqnarray}
&&\ddot{\phi}+3{\dot{a} \over a}\dot{\phi}+m^2\phi =0 \, \\
&&{\ddot{a} \over a}+2\dot{\phi}^2-m^2\phi^2=0 \, .
\end{eqnarray}
Here an overdot denotes $N^{-1}d/dt$.
Variation of the lapse $N$ leads to the constraint $H=0$.

The quantum description employs the ``wavefunction of the
universe'' $\Psi(a,\phi)$, which obeys the Wheeler-DeWitt
equation $\widehat{H}\Psi=0$. Adopting a particular factor ordering
and changing coordinates to $u=a e^{-\phi}$ and $v= a e^{\phi}$, we
find
\begin{equation}\label{wd}
\left( 4 {\partial \over \partial u}{\partial \over \partial v}
+V(u,v)\right)\Psi(u,v)=0 \, ,
\end{equation}
where the minisuperspace potential now reads
\begin{equation}
V(u,v)= uv\left( {m^2 \over 4}uv\left(\ln(v/u)\right)^2-1\right) \, .
\end{equation}

In order to solve these equations we need boundary conditions.
Hartle and Hawking have proposed that, in the Euclidean regime, the
universe does not have any boundaries in space or time. The Euclidean
version of the Lorentzian metric (\ref{lorentz}) can be expressed in the
form
\begin{equation}
ds^2=d\tau^2+a^2(\tau) d\Omega_{(3)}^2 \, .
\end{equation}
The no-boundary proposal demands that the geometry be
compact and the matter fields regular. These conditions
restrict the boundary conditions at zero volume $(\tau=0)$ to be
\begin{equation}
a=0\, ,\quad {d a \over d \tau}=0 \, , \quad \phi=\phi_0\, ,
\quad {d \phi \over d \tau}=0 \, .
\end{equation}
By integrating the Euclidean action from $\tau=0$ to a nearby point
$(a,\phi)$ for large $\phi_0$ and applying WKB matching techniques,
Hawking\cite{hawk3} obtained the boundary condition $\Psi=1$ along
the characteristics $u=0$ and $v=0$. We will subsequently question
the validity of this procedure. The corresponding classical solutions
take the form
\begin{equation}
a \approx (m^2\phi^2)^{-1/2}\sin((m^2\phi^2)^{1/2}\tau)\, , \quad
\phi \approx \phi_0 \, ,
\end{equation}
with $\phi_0$ a constant. The Euclidean solution is matched onto
the Lorentzian metric (\ref{lorentz}) near a totally geodesic
spacelike surface $\Sigma=\partial S^4=S^3$\cite{hg} where $d a/d\tau=0$.
The rotation to imaginary time is thus accomplished by taking
$\tau=\pi/(2m\phi)+it$, leading to the Lorentzian boundary conditions
\begin{equation}
a_{0}\approx{1 \over m\phi_{0}}\, , \quad \dot{a}_{0}\approx 0\, ,
\quad \phi=\phi_{0}\, ,\quad \dot{\phi}_{0}\approx 0 \, .
\end{equation}

Typical Lorentzian solutions see the universe inflate,
then enter into a dust filled state where $\phi$
oscillates rapidly, before recollapsing to a final singularity
as $\phi \rightarrow \pm \infty$ monotonically. In addition to these
typical solutions there are an infinite collection of anomalous
solutions\cite{hawk2}.
The anomalous solutions expand, then recollapse, then bounce and expand
again. The cycle of expansion and collapse may continue indefinitely,
or may terminate after a finite number of bounces. One of these
bouncing solutions is shown in Fig.~1. Also marked are the lines
$V\approx 0$ where the bounces occur, and the line $uv=1$. It is
difficult to track bouncing solutions for more than three of four
bounces as they are exquisitely unstable.

\begin{figure}[h]
\vspace{55mm}
\includegraphics{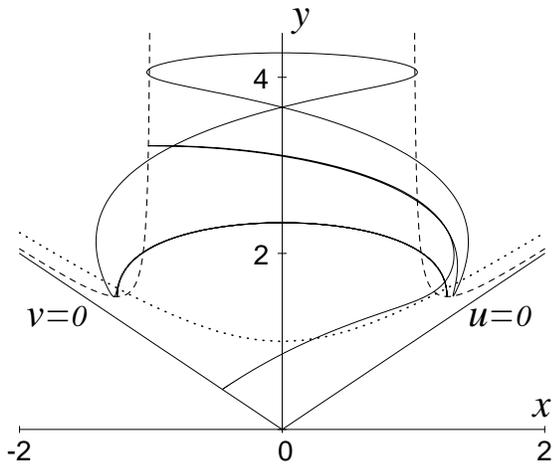}
\vspace{8mm}
\caption{A bouncing classical trajectory in the $(x,y)$ plane. Also
shown are the lines $V(u,v)=0$ (dashed) and $uv=1$ (dotted).
Here $x=(v-u)/2$ and $y=(v+u)/2$.}
\end{figure}

In a remarkable paper, Page\cite{page} conjectured that the anomalous
solutions comprise a fractal set of perpetually bouncing universes.
What Page described is now known as a strange saddle
or strange repellor\cite{wid} -- the analog of a
strange attractor for non-dissipative chaotic systems.

The unstable periodic orbits partition the space of initial conditions
(here just the value of $\phi_0$) according to their outcome. Unless a
universe bounces perpetually it must eventually encounter one of the
cosmological singularities at $a=0$, $\phi=\pm \infty$, corresponding
to the lines $u=0$, $v=0$ respectively. We can confirm Page's
conjecture by studying the boundaries between the two attractors $u=0$
and $v=0$. We do this by color coding $\phi_0$ according to the
outcome; grey for $u=0$ and black for $v=0$. 
The attractor basin boundaries are shown in Fig.~2 for a
representative range of $\phi_0$. Each successive strip of ``Universe DNA''
resolves a small portion of the previous strip. We were able to track
the fractal structure over 12 decades in magnification before
saturating machine precision.

\begin{figure}[h]
\vspace{55mm}

\includegraphics{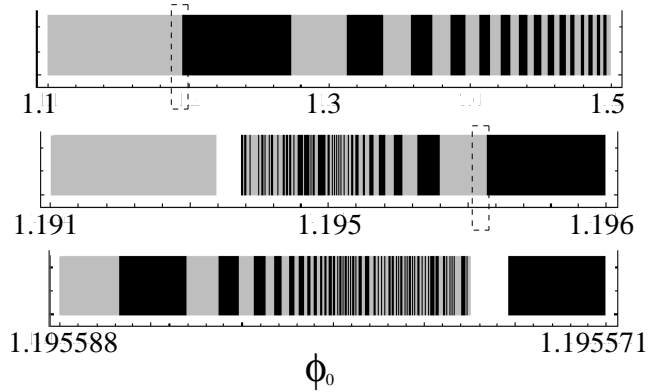}
\vspace{5mm}
\caption{``Universe DNA'': Zooming in on a boundary.}
\end{figure}
 
In the close up views we start to see a third,
white outcome occuring. These regions correspond to trajectories that inflate
by at least 10 $e$-folds to produce a ``big'' universe. A higher
cut-off, such as the cosmologically
interesting value of 60 $e$-folds, would produce much the same pictures,
but with a smaller white region. The lower cut-off was chosen to
keep the numerical integration short. It is interesting that trajectories
near the bouncing solutions can lead to successful inflation (in the
sense of solving the age and flatness problems) without requiring a large
initial value for $\phi$. For our choice of $m=1$, a universe that
never bounces requires $\phi_0 >5.88$ to successfully inflate, while a
universe that bounces just once can successfully inflate if
$\phi_{0}\simeq 1.19$. Although the fractal set of perpetually bouncing
solutions are countable and thus of zero measure, the collection
of universes that bounce at least once is uncountably infinite and of
non-zero measure. For example, if we randomly chose an initial value
of $\phi$ from the region shown in the upper strip of Fig.~2, a ``big''
universe is formed about once in every thousand attempts.

The set of unstable periodic solutions can be quantified by its
spectrum of multifractal dimensions or by its topological
entropy\cite{ott}. Both of these methods provide a coordinate
invariant measure of chaos in general relativity\cite{us}. Here we
calculate the topological entropy as it can be found
analytically. The topological entropy measures the growth in the
number of periodic orbits as their period increases. We quantify the
length of an orbit by the number of symbols needed to describe it.
To do this we need to introduce a symbolic coding. The most efficient
coding we could come up with records the symbol $A$ for each upward
crossing of the line $uv=1$, and the symbol $B$ for each
crossing of the $y$-axis. For example, the bouncing trajectory shown
in Fig.~1 has the coding $ABABABBBABB$.
Applying this recipe to the first four
primary orbits shown in Fig.~3 we obtain the codings
\begin{eqnarray}
{\rm I}=\overline{AB}\, ,& \quad &{\rm II}=\overline{ABB}\, , \nonumber \\
{\rm III}=\overline{ABBB}\, ,&  &{\rm IV}=\overline{ABBBB}\, .
\end{eqnarray}
Here the overline denotes a sequence to be repeated. We can develop a
recurrence relation for the number of periodic orbits, $N(k)$, with
period $k$. Writing $N(k)=P(k)+Q(k)$, where $P(k)$ is the number of
period $k$ words ending in $A$ and $Q(k)$ the number ending in $B$,
we find
\begin{equation}
Q(k+1)=P(k)+Q(k) \, , \quad P(k+1)=Q(k) \, ,
\end{equation}
and $N(2)=2$. The solution is then
$N(k)=(\gamma^{k+1}-\gamma^{-k-1})/\sqrt{5}$, where
$\gamma=(1+\sqrt{5})/2$ is the golden mean. The uncountably infinite
set of aperiodic orbits described by Page correspond to orbits with
$k=\infty$.

\begin{figure}[h]
\vspace{60mm}
\includegraphics{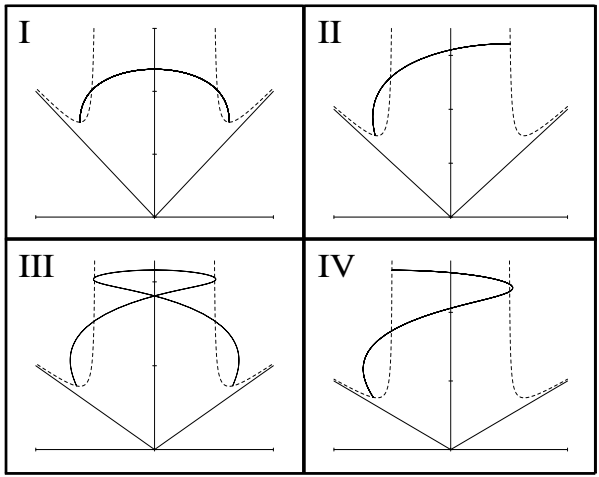}
\vspace{5mm}
\caption{The first four primary orbits in the $(x,y)$ plane.}
\end{figure}

Denoting the number of
periodic orbits with lengths less than or equal to $k$ by ${\cal
N}(k)$, we find the strange repellor has a topological entropy of
\begin{equation}
H_T=\lim_{k\rightarrow \infty}{1 \over k} \ln {\cal N}(k)= \ln \gamma \, .
\end{equation}
Since $H_T>0$, we can conclude that the dynamics is chaotic. It is
interesting to note that the system has a 4-dimensional phase space --
the minimum number required for chaos to occur in a Hamiltonian system.

Having established that the classical evolution is chaotic, we can
consider possible implications for the quantum
theory\cite{berry,rugh}. It has been
suggested that the wave function of the universe will develop small
scale structure if the classical dynamics is chaotic\cite{cal}. This
was demonstrated in Ref.~\cite{cal} by artificially choosing a
potential with an infinite number of discontinuities. In the present
more realistic model, we have to look elsewhere for the source of
fine structure since the potential is completely smooth. Moreover,
if we accept the boundary condition $\Psi=1$, there is no reason to
expect small scale structure to develop since the evolution is
governed by a linear wave equation with a smooth
potential\cite{john}.

On the other hand, it is clear that the WKB approximation breaks down
for our model. This breakdown was independently observed by
Shellard\cite{shell} and Kiefer\cite{kiefer}.
The WKB approximation seeks to interpret $\Psi$ in
terms of classical trajectories according to the decomposition\cite{hp}
\begin{equation}\label{wkb}
\Psi=\Psi_+ +\Psi_-\, ,  \quad\Psi_\pm =\sum_n C_n \exp(\pm i S_n) \, ,
\end{equation}
where each phase factor $S_n$ is taken to obey the Hamilton-Jacobi
equation $(\nabla S)^2 + V =0$. The integral curves of $\nabla S$
correspond to classical Lorentzian solutions:
\begin{equation}
{\partial S \over \partial a}= \pi_a = -{a \over N} \dot{a} \, , \quad
{\partial S \over \partial \phi}= \pi_\phi = {a^3 \over N} \dot{\phi}
\, .
\end{equation}
The WKB approximation is valid if the amplitude $C$ varies much more
slowly than the phase $S$. 
There is a conserved current $J^{\alpha}_n=|C^2_n|
\partial^{\alpha} S_n$ associated with the flux of each WKB wave
packet. The flux of a bundle of classical trajectories
$F_n=\int J_n^{\alpha}\epsilon_{\alpha\beta}dx^{\beta}$ remains constant,
regardless of the hypersurface on which it is sampled\cite{hp}.

In order for the flux to be constant, $|C_n|^2$ must increase if the
bundle of trajectories focuses, and decrease if the bundle defocuses.
When classical trajectories cross, a caustic occurs and
$|C_n|^2\rightarrow \infty$. The basic WKB approximation breaks down
at a caustic, but an enhanced version continues to hold if
the caustic is taken to be the source of a new complex conjugate pair
of WKB solutions\cite{hp}. However, even this enhanced WKB
approximation breaks down when the system is chaotic as regions
near the unstable periodic orbits contain an infinite fractal set of
caustics. The sum over the $\Psi_\pm$ pairs spawned
in these regions does not converge. In addition to the fractal
set of caustics, the strange
repellor also produces rapid defocusing of nearby trajectories. This rapid
divergence drives $|C_n|^2$ rapidly to zero, and the WKB approximation
again fails. Both of these effects can be seen in the numerical studies
of Refs.~\cite{shell,kiefer}.

At this juncture we appear to have a contradiction: solutions of the
Wheeler-DeWitt equation with the boundary condition $\Psi=1$ appear
completely benign, despite the classical evolution being chaotic and the
WKB approximation breaking down.

We believe the reason for the apparent contradiction can be traced to
the boundary condition for $\Psi$. In deriving the condition $\Psi=1$,
Hawking only included a small subset of the non-singular bouncing
universes. This was subsequently criticised by Page\cite{don}, who
argued that the recollapsing, singular paths would modify Hawking's
result. Shellard\cite{shell} attempted to calculate the contribution
from these paths using an enhanced WKB approximation, but was thwarted
by the caustics that developed near the unstable periodic orbits.

Knowing that the root of the problem lies in the
chaotic nature of the classical paths, we can suggest a new approach.
We conjecture that the contribution from the perpetually
bouncing solutions, and the singular paths in their vicinity, can be
found by performing a weighted sum over the multiple instantons that
contribute to the bouncing solutions (see Fig.~4). The sum should be
weighted by the instability exponent for each periodic orbit.
The collection of Euclidean instantons that contribute to the strange
repellor can be viewed as a fractal set of spherical ``Russian
Dolls'', with the radius of each 4-sphere given by $1/(m\phi_0)$ where
$\phi_0$ takes all values on the fractal boundary.

\begin{figure}[h]
\vspace{60mm}
\includegraphics{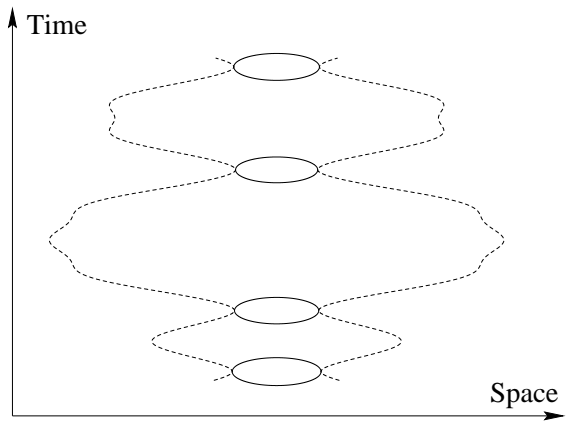}
\vspace{5mm}
\caption{Some of the Euclidean instantons (solid circles) that
contribute to the Lorentzian bouncing solution (dashed line) displayed
in Fig.~1.}
\end{figure}

Our proposal is motivated by Gutzwiller's\cite{gutz} approach to 
quantum chaos. The Gutzwiller trace formula expresses the
quantum-mechanical energy levels of a chaotic system in terms of a sum
over the classical unstable orbits. By adapting some of the
techniques\cite{berryh} developed to evaluate the Gutzwiller sum, we
may be able to properly describe the fractal boundary conditions and
subsequent chaos in the wave function. It would be interesting to
study how these issues impact other approaches to quantum cosmology,
such as Vilenkin's tunneling proposal\cite{alex}. A breakdown in the
WKB approximation would complicate the notion of ``out-going''
wavefunctions used in the tunneling approach.

While our arguments are based on a particular model, in the context of
a particular approach to quantum cosmology, the issues we have raised
will affect any theory of quantum gravity since all dynamical systems
need boundary conditions, and generic dynamical systems are chaotic.

We benefitted from discussions with Janna Levin and John Stewart.

\end{document}